# Efficiency of a thermodynamic motor at maximum power

M. Moreau*, B. Gaveau * and L.S. Schulman**

* University Pierre et Marie Curie, Paris
** Clarkson University

**Abstract.** Several recent theories address the efficiency of a macroscopic thermodynamic motor at maximum power and question the so-called "Curzon-Ahlborn (CA) efficiency". Considering the entropy exchanges and productions in a n-sources motor, we study the maximization of its power and show that the controversies are partly due to some imprecision in the maximization variables. When power is maximized with respect to the system temperatures, these temperatures are proportional to the square root of the corresponding source temperatures, which leads to the CA formula for a bi-thermal motor. On the other hand, when power is maximized with respect to the transitions durations, the Carnot efficiency of a bi-thermal motor admits the CA efficiency as a lower bound, which is attained if the duration of the adiabatic transitions can be neglected. Additionally, we compute the energetic efficiency, or "sustainable efficiency", which can be defined for n sources, and we show that it has no other universal upper bound than 1, but that in certain situations, favorable for power production, it does not exceed ½.

## 1. Introduction.

The efficiency $\gamma_C$ of a thermal motor was defined by Carnot as the ratio of the power produced over the heat received from the higher temperature source. Carnot efficiency played a crucial role in theoretical and applied thermodynamics, which especially distinguished its upper bound $\eta_C$, the celebrated Carnot limit $\eta_C = 1-T_2/T_1$ , $T_1$ and $T_2$ being the temperatures of the hot and cold sources. This upper bound can only be attained when reversibility is realized so that the power produced vanishes. Thus the Carnot limit is not appropriate for an actual motor, which must have a finite power production. Many authors over the years [2-17] have considered the Carnot efficiency of a bi-thermal motor when it produces its maximum power either for macroscopic motors, or for microscopic systems. Clearly, this maximum depends on the parameters which are supposed to be varied, and different responses can be found in different conditions. Nevertheless, the efficiency at maximum power has be found by many authors [2-7] to be the so-called Curzon-Ahlborn (CA) value $\eta_{CA} = 1-(T_2/T_1)^{1/2}$. The hypotheses used for deriving this formula became more and more sophisticated, and it was extended to broader situations [18-21]. Recently, a new energetic efficiency, called the "sustainable efficiency" $\gamma_S$ , was proposed for stationary systems in the framework of stochastic thermodynamics [13-15]. According to this definition, which applies for an arbitrary number of temperature sources, $\gamma_S$ is the ratio of the power produced over the maximum power which could be produced if the power dissipation could vanish. It can be shown that, under specific but reasonably wide conditions, its maximum value is ½ : this conclusion implies that Carnot efficiency at maximum power may be higher than the Curzon-Ahlborn value $\eta_{CA}$ for the stochastic motors described by this theory. The same conclusion was obtained shortly afterwards for classical thermodynamic motors [16] thanks to very general arguments.

The present work is devoted to macroscopic thermodynamic motors, including an arbitrary number of sources, in the context of endoreversible systems [8,22]: the entropy production is due to heat exchanges only, all other sources of entropy creation being neglected. We study their efficiency at maximum power by computing the entropy exchanges and productions in each transition of a cycle, after discussing the assumptions currently used in similar studies, and avoiding some questionable hypotheses used in them. We carefully distinguish different kinds of power maximizations, showing that when the power is maximized with respect to the system temperatures (different from the source temperatures) the Carnot efficiency of a bi-thermal motor has the CA value $\eta_{CA}$. On the contrary, if power is maximized with respect to the durations of the transitions, $\eta_{CA}$ is a lower bound of the efficiency, which can in principle attain the Carnot limit. Additionally, the energetic efficiency previously named "sustainable efficiency" for stochastic systems [15], is defined and computed for n-sources macroscopic motors. This quantity proves to have properties similar to those of its stochastic version, but with different consequences.

Before addressing these points, it is useful to discuss the pioneering derivations of the Curzon-Ahlborn bound [2-4], which used the simplest and perhaps clearest, although imperfect, method: this is the pur pose of Section **2**. In Section **3**, we define a generalized cyclic motor, compute its power production and maximize it with respect to the system temperatures. The maximization with respect to the transition durations is addressed and discussed in Section **4**. Eventually, the sustainable efficiency of macroscopic motors is studied in Section **5**.

## 2. A short discussion of the earliest derivations of the Curzon-Ahlborn bound.

The original derivation of Carnot efficiency at maximum power by Yvon [2] is not known in all details, since this author only sketched it in a lecture at a Genova Conference in 1955 and, as far as we know, did not published it completely elsewhere. Nevertheless, his reasoning was clearly described and apparently very simple. In the next years, Chambadal [2] and Novikov [3] followed the same method, which can be summarized as follows. The engine performs a Carnot cycle between the hot and cold sources at respective temperatures $T_1$ and $T_2$, but the actual temperature of the system during its contact with the hot source should be $T'_1 < T_1$ in order to have a finite heat flux input $\dot{Q}_1 = K_1(T_1 - T'_1)$, where $K_1$ is a constant including the thermal conductivities and the areas of the walls allowing for the heat exchanges with the hot source. The efficiency of the engine has the Carnot value $(1-T_2/T'_1)$ and the power produced is $P = K_1(T_1 - T'_1)(1 - T_2/T'_1)$. It is then straightforwardly found that the power is maximum for $T'_1 = (T_1 T_2)^{1/2}$, which yields the Curzon-Ahlborn value of Carnot efficiency

$$\eta_{CA} = 1 - (T_2/T_1)^{1/2} \tag{1}$$

As mentioned above, this formula was re-derived later by several authors [5-8] from more sophisticated arguments, but the previous method may be the simplest one. Nevertheless, it is clearly imperfect for several reasons, in particular because, logically, one should also consider that the lowest temperature $T'_2$ of the Carnot cycle is higher than the cold source temperature: $T'_2 > T_2$. Furthermore, the durations of the various phases of the Carnot cycle are not taken into account, although they clearly play a role in power production. These points were addressed by further researchers [5-8] in different formalisms, and they can be implemented in the original method, as shown in Appendix A.

The average power produced by the system can be maximized with respect, not to $T'_1$, $T'_2$, but to other variables, such as the durations $\tau_1$, $\tau_2$ of the exchanges with the reservoirs: this was the point of view considered, for instance, in Ref. [16]. It should be noticed that the durations $\tau_1$ and $\tau_2$ generally depend on $T'_1$ and $T'_2$. Assume for instance that the system is a perfect gas performing a Carnot cycle including (*1*) the isothermal expansion ($T'_1$) at temperature $T'_1$ from the initial volume $V_1$ to volume $V_2$, (*2*) the adiabatic expansion ($A_1$) from temperature $T'_1$ and volume $V_1$ to temperature $T'_2$ and volume $V_3$, (*3*) the isothermal expansion ($T'_2$) at temperature $T'_2$ from volume $V_3$ to volume $V_4$, and (*4*) the adiabatic expansion ($A_2$) from temperature $T'_2$ and volume $V_4$ to temperature $T'_1$ and volume $V_1$. Let $\tau_1, \tau_{a1}, \tau_2, \tau_{a2}$ be the respective durations of these successive steps and $\tau$ be the total duration of the cycle. If the molar heat capacities at constant volume, $C_V$, and at constant pressure, $C_P$, are constant it is well known that

$$\frac{V_3}{V_2} = \frac{V_4}{V_1} = \left(\frac{T'_1}{T'_2}\right)^{C_V/R} \equiv \lambda \tag{2}$$

where R is the perfect gas constant and $R/C_V = C_P/C_V - 1 \equiv \gamma - 1$. In the simplest circumstances, the gas is contained in a cylinder closed by a piston moving with the constant velocities v during the expansions and $-v$ during the compressions. Then the ratios $\tau_i/t$ are easily expressed in function of $\lambda$ and of the compression ratio $c = V_2/V_1$. In particular we have

$$\tau_2 = \lambda \tau_1 \tag{3}$$

which shows that the durations of the different steps do depend on the temperatures. Moreover, the ratio of the total time of the adiabatic processes over the total time of the heat exchanges is

$$\frac{\tau_{a1} + \tau_{a2}}{\tau_1 + \tau_2} = \frac{(c+1)(\lambda-1)}{(c-1)(\lambda+1)} \qquad (4)$$

and it also depends on the temperatures, contrary to the assumption used in Ref. [5] for deriving the famous value (1). Of course, the previous relations are not general, but it is clear that the independence of the temperature and the durations is not fulfilled. Even if the relations between the durations and the temperatures can be taken into account, it is hopeless to do so if we look for general results. However, the durations of the different phases of the cycle are much less correlated in the case of an imperfect cycle, which does not return exactly to its initial state. Thus, we will neglect these correlations, as done by all previous authors. Further arguments justifying this procedure will be given later.

Eventually, we remark that it is generally assumed the literature [16] that the weak dissipation regime holds, implying that $\tau_{a1} + \tau_{a2} \ll \tau_1 + \tau_2$. However, formula (4) shows that this is only valid if $\lambda \approx 1$. In practical cases, a typical value of $C_V$ is of the order of 2.5 R and $T_1/T_2$ ranges between 2 and 3 (see Refs.[4,13]). At the optimal regime of Curzon-Ahlborn $T'_1/T'_2 \approx (T_1/T_2)^{1/2} \approx 1.6$, and $\lambda \approx (1.6)^{2.5} \approx 3.1$: then the weak dissipation approximation is questionable.

## 3. Entropy production in a generalized motor.

### 3.1. A general cyclic engine.

Let us consider a system performing a cycle consisting of N, possibly infinitesimal, steps i= 1, 2, …N, with respective positive durations $\delta t_1, \delta t_2,...\delta t_N$. During step i the system receives the heat $\delta Q_i$ from a reservoir at fixed temperature $T_i$ whereas the system is at temperature $T'_i$, which may be considered to be constant if $\delta t_i$ is small enough. $\delta Q_i$ can be 0, and several successive temperatures $T_i$ may be identical whereas the corresponding $T'_i$ could in principle be different, thus describing a time dependent system temperature. Assuming that the laws of irreversible thermodynamics near equilibrium hold, we can write by Fourier law

$$\delta Q_i = K_i (T_i - T'_i) \, \delta t_i \qquad (5)$$

It can be remarked that other expressions of the heat fluxes have been used [22], without any definitive theoretical support. Such expressions are equivalent to the Fourier law for very small temperatures difference between the source and the system, but not if this difference becomes significant, while remaining relatively small. We assume that Fourier law remains valid in this case. In fact, although this classical law is only approximate, its validity has been confirmed both experimentally and theoretically in many circumstances (see for instance, among the abundant literature, [23,24] and references herein).

The adiabatic transitions are considered in the same formalism by taking $K_i= 0$ for them. According to elementary calculations, the entropy change of the reservoir during step i is

$$\delta S_i = \frac{-\delta Q_i}{T_i} \qquad (6)$$

and the entropy change of the system is

$$\delta S'_i = \frac{\delta Q_i}{T'_i} \equiv -\delta S_i + \delta s_i \qquad (7)$$

$-\delta S_i$ being the entropy received from the reservoir and $\delta s_i$ the entropy produced during the exchange, so

$$\delta s_i = \delta Q_i \left( \frac{1}{T'_i} - \frac{1}{T_i} \right) = K_i \frac{(T_i - T'_i)^2}{T_i T'_i} \delta t_i \qquad (8)$$

The work produced during the whole cycle is

$$-W = \sum_i \delta Q_i = K_i (T_i - T'_i) \, \delta t_i \qquad (9)$$

while the total entropy variation of the system vanishes

$$0 = \sum_i \frac{\delta Q_i}{T'_i} = \sum_i K_i \frac{T_i - T'_i}{T'_i} \delta t_i \tag{10}$$

We remark that for these classical formulas to be valid, the system temperature $T'_i$ should be positive, so

$$\delta S'_i = K_i \frac{T_i - T'_i}{T'_i} \delta t_i > - K_i \delta t_i \tag{11}$$

which we will always assume. If now we let $\delta t_i \to 0$, the previous formulas apply, replacing the finite increments by differentials and the sums by time integrals.

### 3.2. Maximization of the power production with respect to the temperatures.

If we assume that the durations of the transitions are fixed whereas the system temperatures $T'_i$ are varied while respecting equality (10), it is easily found that the maximum power per cycle is attained when for each non adiabatic step

$$T'_i = (\lambda T_i)^{1/2} \quad , \text{ with } \quad \lambda^{1/2} = \frac{\sum_i K_i (T_i)^{1/2} \delta t_i}{\sum_i K_i \delta t_i} \tag{12}$$

the maximum power produced being

$$P_{max} = \frac{\left(\sum_i K_i T_i \delta t_i\right)\left(\sum_i K_i \delta t_i\right) - \left(\sum_i K_i (T_i)^{1/2} \delta t_i\right)^2}{\left(\sum_i K_i \delta t_i\right)\left(\sum_i \delta t_i\right)} \geq 0 \tag{13}$$

In the case of infinitesimal time increments $\delta t_i$, Eqs.(11,12) apply for each non-adiabatic phase, replacing the sums by the corresponding integrals.

Eq.(12) implies that the maximum power is obtained the system temperature $T'_i$ is constant during the heat exchange with any source i at temperature $T_i$. In practice this condition may be difficult to satisfy, and it certainly implies particular mechanisms, but for maximizing the power it is favorable to approach it as far as possible. Thus, as most authors, we now on assume that this condition is realized: $\tau_i \equiv \delta t_i$ will represent the total duration of the exchanges with source i, $\delta_i \equiv \delta S'_i$ being the corresponding variation of the system entropy.

If there are only two reservoirs with temperatures $T_1$ and $T_2$ ($T_1 > T_2$), the heat received from reservoir i (i = 1 or 2) is $\delta Q_i = (T_i - T'_i) K_i \tau_i = T'_i \delta_i$, and Eq. (10) indeed shows that the Carnot efficiency has the CA value $\eta_{CA}$.

This method can be extended to an engine that does not perform an exact cycle. More precisely, it can happen that after N steps, the engine does not return exactly to the initial state, but only to a neighborhood of it. If the state, for instance, is characterized by the position of a piston and the system temperature, it is sufficient in practice that these final values are equal to the initial ones in average in order that the system can reach a permanent regime (provided that it is conveniently supplied with heat!). At the end of each pseudo-cycle, the left-hand sides of Eqs. (A.2,3) should be corrected by adding the respective variations of the internal energy and of the entropy of the system $\Delta S'$ and $\Delta U'$. If the time average of these quantities vanishes on a large number of pseudo cycles, the previous results are expected to hold in average. Thus, as mentioned in Section **2**, the relations that may exist between the durations of the different transitions of an exact cycle, such as relations (3,4), lose most of their meaning in a stochastic version of pseudo-cycles. This is why we can ignore them here.

### 4. Maximization of the power production with respect to the transitions durations.

In the previous multi-temperature machine, let us now suppose that we can vary the durations of the different steps. We first remark that by Eq.(8) the entropy production during step i can be written for a non adiabatic transition

$$\delta s_i = K_i \frac{(T_i - T'_i)^2}{T_i T'_i} \delta t_i = \frac{-\delta S_i \, \delta S'_i}{K_i \, \delta t_i} \tag{14}$$

During an adiabatic transition i, we assume that $\delta S_i = \delta S'_i = \delta s_i = 0$. Using (14) if i is not adiabatic we obtain

$$\delta s_i = \frac{(\delta S'_i)^2}{K_i \, \delta t_i}\left(1 + \frac{\delta S'_i}{K_i \, \delta t_i}\right)^{-1} \tag{15}$$

Writing $\tau = \sum_{i=1,\ldots N} \delta t_i$, the power produced after N steps, *i.e.* after a (pseudo) cycle, is

$$\begin{aligned}P &= \frac{-W}{\tau} = \frac{1}{\tau}\sum_i^* \delta Q_i = \frac{1}{\tau}\sum_i^* T_i(-\delta s_i) \\ &= \frac{1}{\tau}\sum_i^* T_i(\delta S'_i - \delta s_i) = \frac{1}{\tau}\sum_i^* T_i \frac{\delta S'_i}{1 + \frac{\delta S'_i}{K_i \, \delta t_i}}\end{aligned} \tag{16}$$

where $\sum^*$ denotes the sum over non adiabatic transitions. From now on, we will use the previous formulas with the condensed notation $\delta_i \equiv \delta S'_i$ and $\tau_i \equiv \delta t_i$, assuming that, according to Section 3.2., $\delta_i \succ - K_i \tau_i$.

### 4.1. Weak dissipation regime.

The weak dissipation regime (see for instance [16]) can be considered as the standard situation where the usual laws of irreversible thermodynamics hold. Then the heat exchanges are slow and we can consider that $\delta_i \ll K_i \tau_i$. So the power produced is

$$P \approx \frac{1}{\tau}\sum_i^* T_i\left(\delta_i - \frac{(\delta_i)^2}{K_i \tau_i}\right) \tag{17}$$

According to the methods of Ref.[16], we can maximize the power produced with respect to the $\tau_i$, considering that the entropy variation $\delta_i = \delta S'_i$ of the system during each step i is fixed. Such a choice implies that the entropy changes of the system are relevant quantities, to be used in the best possible way: this arbitrary convention is contrary to the usual consideration that only the energy inputs import, but it is reasonable in the scope of sustainable development, now adopted in many circumstances. Then, we obtain for non adiabatic transitions

$$\frac{T_i (\delta S'_i)^2}{K_i (\tau_i)^2} = P \tag{18}$$

where we consider only the situations when P is positive. Thus, if transition i is not adiabatic

$$\tau_i = |\delta_i| P^{-1/2} \left(\frac{T_i}{K_i}\right)^{1/2} \tag{19}$$

whereas the adiabatic transitions should obviously be as short as possible in order to maximize the power produced. The maximum power and the heat received from each source can be expressed (Appendix B) in terms of the $\delta_i$ and of the other parameters.

In the case of a bi-thermal motor operating successively with two sources at temperatures $T_1$ and $T_2$ ($T_1 > T_2$) we obtain (Appendix B)

$$P^{1/2} = \frac{1}{2}(T_1 - T_2)\left[\left(\frac{T_1}{K_1}\right)^{1/2} + \left(\frac{T_2}{K_2}\right)^{1/2}\right]^{-1} \tag{20}$$

and the Carnot efficiency is found to be

$$\lambda_C = 1 + \frac{\delta Q_2}{\delta Q_1} = \frac{T_1 - T_2}{2T_1 - (T_1 - T_2)\left[1 + (T_2/T_1)^{1/2}(K_1/K_2)^{1/2}\right]^{-1}} \qquad (21)$$

We recover the main result of Ref.[16] as well as its consequences. $\lambda_C$ is a decreasing function of $K_1/K_2$: if this ratio tends to $\infty$, one obtains a lower bound of $\lambda_C$

$$\lambda_{C_{min}} = \frac{1}{2}\left(1 - \frac{T_2}{T_1}\right) = \frac{1}{2}\eta_C$$

whereas if $K_1/K_2 \to 0$, one finds the upper bound $\overline{\eta}_C$ which was also obtained for stationary stochastic motors [14,15]

$$\lambda_{C_{max}} = \frac{T_1 - T_2}{T_1 + T_2} = \overline{\eta}_C$$

Eventually, if $K_1/K_2 = 1$, the Curzon-Ahlborn value $\eta_{CA}$ is recovered. We will see, however, that these conclusions are not preserved in a more general regime, when the weak dissipation approximation does not hold.

### 4.2. A generalized regime.

We now maximize the general expression (16) of the power produced with respect to the $\tau_i$, again considering that the entropy variation $\delta_i \equiv \delta S'_i$ of the system during each step i is fixed. Thus, we admit that the Fourier law remains valid outside the weak dissipation regime, when $\delta S'_i$ is not necessarily much smaller than $K_i \tau_i$, which implies that $|T'_i - T_i|$ can be of the order of $T'_i$. In fact, this is currently realized in actual heat engines where, nevertheless, the Fourier law is assumed to be valid. In this case, maximizing expression (15) of P>0 yields, for a non adiabatic transition

$$\frac{T_i(\delta_i)^2}{K_i(\tau_i)^2\left[1 + \frac{\delta_i}{K_i \tau_i}\right]^2} = P \qquad (22)$$

if $\tau_i > 0$. It results from (22) that

$$\frac{\frac{\delta_i}{K_i \tau_i}}{1 + \frac{\delta_i}{K_i \tau_i}} = \varepsilon_i \left(\frac{P}{K_i T_i}\right)^{1/2} \qquad (23)$$

where $\varepsilon_i$ is the sign of the left hand side. We have seen that $\delta_i > -K_i \tau_i$, so that $\varepsilon_i = \text{sign}(\delta_i)$. Thus

$$\tau_i = \frac{|\delta_i|}{K_i}\left[\left(\frac{P}{K_i T_i}\right)^{-1/2} - \varepsilon_i\right] \qquad (24)$$

provided that the right hand side being positive if $P < K_i T_i$ for each step with positive entropy variation. It is easily seen from (16) that this inequality is satisfied for at least one of these steps. If the inequality was not satisfied for some step j with positive entropy variation, this step j should be skipped ($\tau_j = 0$) in order to maximize the power production. We assume that such steps, if any, have been suppressed. On the other hand, it can checked by (24) that $\tau_i$ satisfies inequality (11), as it should be. From (16, 22) and (24) we obtain

$$\begin{aligned}\tau P &= A - P^{1/2} B \\ \tau &= \tau_a + P^{-1/2} B - C\end{aligned} \qquad (25)$$

with

$$A = \sum_{i}^{*} T_i \delta_i, \quad B = \sum_{i}^{*} \left(\frac{T_i}{K_i}\right)^{1/2} |\delta_i|, \quad C = \sum_{i}^{*} \frac{1}{K_i} \delta_i \qquad (26)$$

It is clear that B > 0 and that, by (25), P > 0 implies A > 0 (so, obviously, all the temperatures $T_i$ cannot be equal), but C may be negative as well as positive. Eqs.(25) yield

$$(C - \tau_a) P - 2B P^{1/2} + A = 0 \qquad (27)$$

which has (at least) one positive solution if

$$\tau_a \geq C - B^2/A \qquad (28)$$

It is always possible to satisfy (28) if $\tau_a$ is large enough, but then the power production is small. In order that no minimum value is assigned to $\tau_a$, it is desirable that $B^2 - AC \geq 0$: we assume that we are in this situation. Then Eq.(27) has one positive solution such that $\tau > 0$

$$P^{1/2} = \frac{A}{B + (B^2 - AC + \tau_a A)^{1/2}} \qquad (29)$$

This formula shows that, in order to maximize P, the duration $\tau_a$ of the adiabatic transitions should be as small as possible, as already pointed out. We see that the power production P can be higher than the value $A^2/(2B)^2$ obtained in the standard weak dissipation regime (see Appendix B), provided that $C > \tau_a$. So, in order to consider a favorable situation, we may assume that C is positive and that $\tau_a$, if not completely negligible, is at least less that C: $\tau_a < C$.

The heat exchanged with source i is by (17)

$$\delta Q_i = T_i \frac{\delta_i}{1 + \frac{\delta S_i}{K_i \tau_i}} = T_i \delta S_i - P^{1/2} \varepsilon_i \delta S_i \left(\frac{T_i}{K_i}\right)^{1/2} \qquad (30)$$

and the temperature $T'_i$ of the system during the heat exchanges with source i is

$$T'_i = T_i - P^{1/2} \varepsilon_i \left(\frac{T_i}{K_i}\right)^{1/2} \qquad (31)$$

which implies that $P < K_i T_i$ for each step i such that $\delta S'_i > 0$, as noticed previously. It can be shown that this solution is indeed a maximum of the power production.

As an example, let us consider a bi-thermal motor with sources at temperatures $T_1$ and $T_2$ ($T_1 > T_2$). We have $\delta S'_1 = -\delta S'_1 > 0$, and

$$\delta Q_1 = T_1 \delta S'_1 \left(1 - \frac{P^{1/2}}{(K_1 T_1)^{1/2}}\right), \quad \delta Q_2 = -T_2 \delta S'_1 \left(1 + \frac{P^{1/2}}{(K_2 T_2)^{1/2}}\right)$$

It is found that $B^2 - AC = (T_1/K_2 + T_2/K_1)^2 > 0$, so that $\tau_a$ may indeed be arbitrarily small. (29) yields

$$P^{1/2} = \frac{(T_1 - T_2)}{\left[(T_1/K_1)^{1/2} + (T_2/K_2)^{1/2}\right] + \left[\left((T_1/K_2)^{1/2} + (T_2/K_1)^{1/2}\right)^2 + \tau_a(T_1 - T_2)\right]^{1/2}} \qquad (32)$$

If $\tau_a$ is small enough to be neglected we obtain

$$P^{1/2} \approx \frac{(T_1)^{1/2} - (T_2)^{1/2}}{(K_1)^{-1/2} + (K_2)^{-1/2}} \equiv (P_0)^{1/2} \qquad (33)$$

$P_0^{1/2}$ being, furthermore, an upper bound for $P^{1/2}$. The Carnot efficiency is

$$\gamma_C = 1 + \frac{\delta Q_2}{\delta Q_1} = \frac{(T_1 - T_2) - P^{1/2}\left[(T_1/K_1)^{1/2} + (T_2/K_2)^{1/2}\right]}{T_1 - P^{1/2}(T_1/K_1)^{1/2}} \equiv \gamma_C(P) \tag{34}$$

which is a decreasing function of P. Cumbersome but elementary calculations show that if $\tau_a \to 0$,

$$\gamma_C \to \gamma_C(P_0) = 1 - \left(\frac{T_2}{T_1}\right)^{1/2} = \eta_{CA} \tag{35}$$

Thus, when the total duration of the adiabatic steps tends to 0, the Curzon-Ahlborn efficiency is recovered for all values of $K_1$ and $K_2$, and not only when $K_1 = K_2$ as found in the weak dissipation regime. Moreover, for finite $\tau_a$, $P < P_0$, so that

$$\gamma_C(P) > \gamma_C(P_0) = \eta_{CA} \tag{36}$$

which is one of our main results: the Curzon-Ahlborn efficiency is a lower bound of Carnot efficiency at maximum power. A similar conclusion was obtained, in certain circumstances, from the explicit study of the so-called 3-level model motor [17].

## 5. Multi-sources motor and sustainable efficiency

### 5.1. Sustainable efficiency of a macroscopic thermodynamic motor

The notion of sustainable efficiency was introduced in stochastic thermodynamics for a mesoscopic motor operating in a non-equilibrium stationary state in Ref.[14], noticing that the power produced P can be written

$$P = A - D_P$$

$D_P$ being the power dissipation, *i.e.* the energetic equivalent of the entropy production rate, which is always positive out of equilibrium. A is the power which would be produced if the power dissipation could vanish. The sustainable efficiency $\gamma_S$ was defined [14] as

$$\gamma_S = \frac{P}{A} = \frac{P}{P + D_P}$$

Thus, the sustainable efficiency makes sense for any multisource motor without favoring one the sources, and it may be appropriate for an unbiased estimation of the motor performances. For these reasons, and its relations with Carnot efficiency [14,15] it is interesting to extend this concept to the present macroscopic formalism. The heat $\delta Q_i$ received by the system during step i can be written by (15)

$$\delta Q_i = -T_i \delta S_i = T_i \delta S'_i - T_i \delta s_i$$

where $\delta S_i$ and $\delta S'_i$ are the entropy variations of reservoir i and of the system, respectively, and $\delta s_i$ is the entropy production during step i. The work produced during the cycle is

$$-W = \sum_i \delta Q_i = \sum_i T_i \delta S'_i - \sum_i T_i \delta s_i \tag{37}$$

The energy dissipation during step i is $T_i \delta s_i$ and the total energy dissipation during the cycle is

$$D_W = \sum_i T_i \delta s_i \tag{38}$$

whereas $A \equiv \sum_i T_i \delta S'_i = -W + D_W$ is the maximum work that could be produced during a cycle if all dissipation could be avoided. In fact, during step i the temperature of the heat source being $T_i$, and the exergy of the system (in the absence of a pressure reservoir) is $Ex_i = E'_i - T_i S'_i$, $E'_i$ and $S'_i$ being the internal energy and the entropy of the system, respectively. According to classical engineering thermodynamics, the maximum work that can be produced by the system during step i is $-\delta Ex_i$, and the maximum work produced during a complete cycle is $-\sum_i \delta Ex_i = \sum_i T_i \delta S'_i = A$, in agreement with definitions (26). Thus, for a macroscopic engine the sustainable efficiency can be defined by

$$\gamma_S = \frac{-W}{-W+D_W} = \frac{\sum_i \delta Q_i}{\sum_i T_i \delta S'_i} \qquad (39)$$

In similarity with the stochastic case, it is interesting to consider the values of the sustainable efficiency at maximum power. In this situation we use formulas (29,30) and (39) and obtain

$$\gamma_S = \frac{A - P^{1/2} B}{A} = 1 - \frac{B}{B + (B^2 - AC + \tau_a A)^{1/2}} \qquad (40)$$

It is clear that if $\tau_a \leq C$, and in particular if $C > 0$ and $\tau_a \to 0$, $\gamma_S \leq \frac{1}{2}$. This situation is not general: the sustainable efficiency, as defined above, can be larger than ½ at maximum power, and it even tends to 1 if the adiabatic transitions are infinitely slow, as shown by (40): then by (41) the Carnot efficiency reaches the classical Carnot limit, but the power produced vanishes. On the other hand, we have seen above that in order to maximize the power production, it is desirable to minimize the duration of the adiabatic phases and to have a positive C coefficient larger that $\tau_a$ : in such situations, the maximum value of the thermodynamic sustainable efficiency is ½. This upper bound is attained if $C = \tau_a$ and, in particular, when all Fourier coefficients $K_i$ are equal and $\tau_a \to 0$.

It can be remarked that the stochastic sustainable efficiency defined in Refs.[14,15] also admits the upper bound ½ in certain situations, essentially, when the stochastic dynamics is varied while maintaining constant the stationary probability distribution. In the absence of any specific condition, however, the stochastic sustainable efficiency has no general upper bound lower that 1, like its macroscopic version.

**5.2. Macroscopic and stochastic sustainable efficiencies**

At this point, it can be useful to summarize the analogies and differences between these concepts. Clearly, the macroscopic sustainable efficiency considered in Section **5.1.** has much similarity with the stochastic sustainable efficiency of Refs.[14,15], by its definition and it properties. Nevertheless, an important difference between these efficiencies is that the first one only depends on a few macroscopic parameters, whereas the stochastic efficiency depends on the complete dynamics of the system and implies a large number of microscopic parameters. For this reason, the constraint of a constant stationary probability, imposed in maximizing the stochastic sustainable efficiency, can hardly be transposed to the macroscopic case.

These analogies and differences are also manifested in the relation existing between the Carnot and sustainable efficiencies. It can be shown (see Appendix C) that the macroscopic sustainable efficiency satisfies the relation already found [15] in the stochastic case, which reads (if the only sources of entropy production are the heat exchanges, as assumed here)

$$\gamma_C = \left(1 - \frac{T_2}{T_1}\right)\left[1 - T_2 \frac{\delta s_1 + \delta s_2}{(1-\gamma_S)^{-1}(T_1 \delta s_1 + T_2 \delta s_2) - (T_1 - T_2)\delta s_1}\right] \qquad (41)$$

Thus, $\gamma_C$ is an increasing function of $\gamma_S$. It tends to the Carnot limit $\eta_C$ if $\gamma_S \to 1$, which is in principle possible. In practice, we saw that in favorable situations we should have $\gamma_S \leq \frac{1}{2}$. In the macroscopic theory, however this inequality, combined with (41), does not imply any new effective upper bound for $\gamma_C$, because the parameters in (41) cannot been varied independently, as discussed in Appendix C. In the stochastic case, on the contrary, Eq.(41) implies that $\gamma_C \leq (T_1 - T_2)/(T_1 + T_2)$ if the inequality $\gamma_S \leq \frac{1}{2}$ holds [15], whereas in different situations $\gamma_C$ can approach the Carnot limit: this point was illustrated by the complete analytical study of the 3-level stochastic motor [17], and has been developed more generally in a recent article [25].

## 6. Conclusion

The efficiency of a thermal motor in the conditions of maximal power production has been discussed intensively from more than fifty years. The original derivations were completed and generalized, and alternative derivations were proposed, until recent papers contested these apparently well established results, either proposing a higher upper bound of the efficiency, or even asserting that there is no general upper bound other than the Carnot value. These discussions concern the classical engine considered in macroscopic thermodynamics as well as the mesoscopic motors introduced in stochastic thermodynamics. In the present paper, focused on macroscopic thermodynamics, we have shown that the controversies are partly due to the imprecise definition of the maximization conditions. When the maximization is taken over the system temperatures during the non adiabatic transitions (or equivalently over the corresponding system entropy variations) the Curzon-Ahlborn efficiency is obtained for a bi-thermal motor. On the other hand, if the maximization is taken over the durations of the transitions, the CA efficiency is a lower bound of the actual efficiency: it is attained if the total duration of the adiabatic transitions can be neglected with respect to duration of the other transitions.

In analogy with a definition given for stochastic motors, we have also introduced the notion of "sustainable efficiency" to macroscopic motors, *i.e.* the ratio of the power produced to the maximum power which could be produced with the same resources if all irreversible effects could be suppressed. This efficiency $\gamma_S$, not only can be used for any number of reservoirs without favoring one of them, but also gives a new light on the more usual Carnot efficiency. The only general upper bound of the sustainable efficiency is 1, implying that in the most general situation, the Carnot efficiency at maximum power can in principle approach the Carnot limit. Practically, however, maximizing the power is preferably obtained in situations where $\gamma_S \leq \frac{1}{2}$, and the energy exchanges with the reservoir obey the law of heat diffusion: these conditions imply that the Carnot efficiency of a bi-thermal motor has the Curzon-Ahlborn value, as found by direct calculations.

As a conclusion, the concept of efficiency at maximum power can be misleading, because it depends on the kind of maximization which is considered. As well as the other values that have been proposed, the Curzon-Ahlborn efficiency is not "universal". However, it plays a crucial role when the energy exchanges with the reservoirs are governed by the Fourier law: in certain situations, it represents the lowest bound of the efficiency at maximum power, which is attained in specific circumstances. This is why it remains an important value in the theory of macroscopic motors.

## Appendix A. Complements on the earliest derivation of Curzon-Ahlborn bound.

These derivations can be completed to include ingredients which were discarded initially, mainly, the durations of the exchanges. These durations were considered, for instance in Ref[5], with questionable assumptions on the durations of the adiabatic transitions which can be avoided. With the notations of Section 2, we again study a bi-thermal motor operating with a hot source at temperature $T_1$ and a cold source at temperature $T_2$. Assume that during the exchanges with source i (i= 1,2), the heat flux is finite and given by $\dot{Q}_i = K_i(T_i - T'_i) < 0$, $K_i$ being a constant. We now take into account the durations of the different transitions occurring during a Carnot cycle. Let the isothermal phase ($T'_1$) of exchange with the hot source 1 have a duration $\tau_1$, and the isothermal exchange phase ($T'_2$) with the cold source 2 have a duration $\tau_2$, whereas the total duration of the cycle is $\tau$. Considering a stationary statistical ensemble of identical, independent systems operating according to the same Carnot cycle, at a given time the proportion of systems undergoing a transition ($T'_i$) is $\tau_i/\tau$, so that the ensemble average of the heat flux received by a system from source (i) is

$$<\dot{Q}_i> = K_i(T_i - T'_i)\, \tau_i/\tau \equiv \kappa_i(T_i - T'_i) \tag{A.1}$$

which is clearly identical to the time average of this heat flux over a period much larger than $\tau$ for a unique system.

It results from classical thermodynamics that the entropy of the system should not change during a whole cycle so that, if the transitions connecting the isothermal phases are rigorously adiabatic

$$\frac{<\dot{Q}_1>}{T'_1} + \frac{<\dot{Q}_2>}{T'_2} = 0 \tag{A.2}$$

The average work $<\dot{W}>$ received by the system by unit time results from the energy conservation

$$<\dot{W}> + <\dot{Q}_1> + <\dot{Q}_2> = 0 \tag{A.3}$$

Thus, the average power produced by the system is $<P> \equiv -<\dot{W}> = <\dot{Q}_1> + <\dot{Q}_2>$ and its Carnot efficiency is

$$\gamma_C \equiv \frac{<P>}{<\dot{Q}_1>} = 1 - \frac{<\dot{Q}_2>}{<\dot{Q}_1>} = 1 - \frac{T'_2}{T'_1} \tag{A.4}$$

The temperatures $T'_1$ and $T'_2$ cannot be chosen independently, since they must satisfy the relation resulting from (A.1,2)

$$\frac{\kappa_1(T_1 - T'_1)}{T'_1} + \frac{\kappa_2(T_2 - T'_2)}{T'_2} = 0 \tag{A.5}$$

where (A.2,3) obviously imply that $T'_1 < T_1$ and $T'_2 > T_2$. It is easily shown that, if the temperatures $T'_1$ and $T'_2$ are varied while respecting condition (A.2), the power produced $<P>$ is maximum when

$$T'_1 = \left(\alpha_1 \sqrt{T_1} + \alpha_2 \sqrt{T_2}\right)\sqrt{T_1} \quad , \quad T'_2 = \left(\alpha_1 \sqrt{T_1} + \alpha_2 \sqrt{T_2}\right)\sqrt{T_2} \tag{A.6}$$

with $\alpha_i = \frac{\kappa_i}{\kappa_1 + \kappa_2}$ $(i = 1, 2)$. It results from (A.3,4) that the Carnot efficiency has the expected CA value

$$\eta_{CA} = 1 - \sqrt{\frac{T_2}{T_1}} \tag{A.7}$$

**Appendix B. Power maximization in the weak dissipation approximation.**

In the weak dissipation regime, the transitions are slow [16] and we have $\delta_i \ll K_i \tau_i$. Then the power produced is

$$P \approx \frac{1}{\tau} \sum_i^* T_i \left(\delta_i - \frac{(\delta_i)^2}{K_i \tau_i}\right) \tag{B.1}$$

Following the methods of Ref.[16], we now maximize the power produced with respect to the $\tau_i$ considering that the entropy variation $\delta_i = \delta S'_i$ of the system during each step i is fixed. To be physically meaningful, this assumption implies that the important quantities to consider when running a motor are the entropy inputs from or to the reservoirs, rather than the energy inputs. This is not the usual point of view, which focuses on the fuel consumptions or energies rejected to the environment. Nevertheless, we think that the entropy changes could be more significant than the energy variations, since energy is in principle conserved, even if it can hardly be used in certain forms, whereas entropy is not. In any case, this maximization condition can be considered as a mathematical condition but, once more, it is connected with not purely scientific considerations. Then the $\delta_i$ should satisfy the constraint

$$\sum_i \delta_i = 0 \tag{B.2}$$

which, nevertheless, does not affect the variables $\tau_i$. Maximizing P gives for non adiabatic transitions

$$\frac{T_i (\delta S'_i)^2}{K_i (\tau_i)^2} = P \tag{B.3}$$

where we consider only the situations when P is positive. Thus, if transition i is not adiabatic

$$\tau_i = |\delta_i| p^{-1/2} \left(\frac{T_i}{K_i}\right)^{1/2} \qquad (B.4)$$

whereas the adiabatic transitions should obviously be as short as possible in order to maximize the power produced. So

$$P = \frac{1}{\tau}\sum_i^* T_i \left(\delta_i - \frac{(\delta_i)^2}{K_i \tau_i}\right) = \frac{1}{\tau} A - \frac{p^{1/2}}{\tau} B \qquad (B.5)$$

where we have defined

$$A = \sum_i T_i \delta_i \; , \; B = \sum_i^* \left(\frac{T_i}{K_i}\right)^{1/2} |\delta_i| \qquad (B.6)$$

On the other hand if $\tau_a$ is the total duration of the adiabatic steps we have

$$\tau = \tau_a + \tau_b \quad \text{where} \quad \tau_b = \sum_i^* \tau_i = P^{-1/2} B \qquad (B.7)$$

In the weak dissipation approximation $\tau_a$ should be small with respect to $\tau$: $\tau_a \ll \tau$. So, combining (B.5) and (B.7) yields

$$P^{1/2} = \frac{A}{2B}, \quad \tau = \frac{2B^2}{A} \qquad (B.8)$$

The heat received from source (i) at temperature $T_i$ during step i is, in the present approximation

$$\delta Q_i \approx T_i \left(\delta_i - \frac{(\delta_i)^2}{K_i \tau_i}\right) \qquad (B.9)$$

In the case of a bi-thermal motor operating successively with two sources at temperatures $T_1$ and $T_2$ ($T_1 > T_2$)

$$P^{1/2} = \frac{1}{2}(T_1 - T_2)\left[\left(\frac{T_1}{K_1}\right)^{1/2} + \left(\frac{T_2}{K_2}\right)^{1/2}\right]^{-1} \qquad (B.6)$$

and the Carnot efficiency is

$$\gamma_C = 1 + \frac{\delta Q_2}{\delta Q_1} = \frac{T_1 - T_2}{2T_1 - (T_1 - T_2)\left[1 + (T_2/T_1)^{1/2}(K_1/K_2)^{1/2}\right]^{-1}} \qquad (B.7)$$

We recover, with different notations, the main result of Ref.[13], as well as the following conclusions. $\gamma_C$ is a decreasing function of $K_1/K_2$. Its lower bound is obtained if $K_1/K_2 \to \infty$

$$\gamma_{C_{min}} = \frac{1}{2}\left(1 - \frac{T_2}{T_1}\right) = \frac{1}{2}\eta_C$$

whereas if $K_1/K_2 \to 0$, one finds the upper bound $\overline{\eta}_C$ which was also obtained for stationary stochastic motors [14,15]

$$\gamma_{C_{max}} = \frac{T_1 - T_2}{T_1 + T_2} \equiv \overline{\eta}_C$$

Eventually, if $K_1/K_2 = 1$, the Curzon-Ahlborn value $\eta_{CA}$ is recovered. These conclusions, however, essentially depend on the weak dissipation assumption, as shown in Section **4.2**.

## Appendix C. Carnot efficiency and sustainable efficiency for a bi-thermal motor

The Carnot efficiency $\gamma_C$ of a bi-thermal motor exchanging heat with a hot source at temperature $T_1$ and a cold source at temperature $T_2$, as considered in Section 3, can be expressed in terms of its sustainable efficiency $\gamma_S$ defined by (39), as done in Ref.[14,15] for the stochastic efficiency. Using (36) we obtain (40)

$$\gamma_C = \left(1 - \frac{T_2}{T_1}\right)\left[1 - T_2 \frac{\delta s_1 + \delta s_2}{(1-\gamma_S)^{-1}(T_1 \delta s_1 + T_2 \delta s_2) - (T_1 - T_2)\delta s_1}\right] \quad (C.1)$$

where we supposed that there is rigorously no entropy production during the adiabatic phases nor during the energy exchanges with mechanical systems (whereas the contrary hypothesis was considered in the stochastic version of this problem [15]). Then, $\gamma_C$ is an increasing function of $\gamma_S$ if the other parameters remain constant. If $\gamma_S \leq \frac{1}{2}$, as discussed above, we have

$$\gamma_C \leq [1-(T_2/T_1)]\left[1 - T_2 \frac{1+\delta s_2/\delta s_1}{2(T_1 + T_2 \delta s_2/\delta s_1) - (T_1 - T_2)}\right] \quad (C.2)$$

As in the case of stochastic efficiency [15], the right hand side of (C.2) increases from $\frac{1}{2}(1-T_2/T_1)$ to $\bar{\eta}_C = (T_1 - T_2)/(T_1 + T_2)$ when $\delta s_2/\delta s_1$ decreases from $\infty$ to 0, but now $\bar{\eta}_C$ is no effective upper bound of $\gamma_S$, since in practice $\delta s_2/\delta s_1$ cannot vanish, nor vary independently of the other parameters. In fact, as an example, let us consider the symmetric case, with the coefficients $K_1$ and $K_2$ of the Fourier law (5) are equal

$$K_1 = K_2 \equiv K$$

Then it is seen from (26) and (40) that $\gamma_S = 1/2$ (if the duration of adiabatic phases is neglected). From (15) and the maximization condition (23) it is found that $\delta s_2/\delta s_1 = (T_1/T_2)^{1/2}$, and (C.1) implies that

$$\gamma_C = 1 - (T_2/T_1)^{1/2} = \eta_{CA}$$

in agreement with (35).

This result is also valid for a stochastic motor if $\gamma_S \leq \frac{1}{2}$ and if $\delta s_2/\delta s_1 = (T_1/T_2)^{1/2}$. The last relation holds whenever the heat exchanges are governed by diffusion phenomena and obey the Fourier law, but it is not necessarily true if the energy is transferred differently, as it may happen in molecular motors or in complex biochemical phenomena.